\documentclass[prl,aps,twocolumn,showpacs,preprintnumbers,amsmath,amssymb]{revtex4}
\usepackage{dcolumn}
\usepackage{graphicx}

\begin{document}

\title{Correlated Electronic Structures and the Phase Diagram of
Hydrocarbon-based Superconductors}
\author{Minjae Kim$^1$}
\author{Hong Chul Choi$^2$}
\author{Ji Hoon Shim$^{2,1}$}
\email[E-mail: ]{jhshim@postech.ac.kr}
\author{B. I. Min$^1$}
\email[E-mail: ]{bimin@postech.ac.kr}
\affiliation{$^1$Department of Physics, PCTP,
Pohang University of Science and Technology, Pohang, 790-784, Korea}
\affiliation{$^2$Department of Chemistry,
Pohang University of Science and Technology, Pohang, 790-784, Korea}
\date{\today}

\begin{abstract}
We have investigated correlated  electronic structures and the phase diagram
of electron-doped hydrocarbon molecular solids,
based on the dynamical mean-field theory.
We have found that the ground state of hydrocarbon-based
superconductors such as electron-doped picene and coronene is
a multi-band Fermi liquid, while that of non-superconducting
electron-doped pentacene is a single-band Fermi liquid
in the proximity of the metal-insulator transition.
The size of the molecular orbital energy
level splitting plays a key role in producing the superconductivity
of electron-doped hydrocarbon solids. The multi-band nature of
hydrocarbon solids would boost the superconductivity
through the enhanced density of states at the Fermi level.
\end{abstract}


\maketitle

Since the discovery of high $T_C$ superconductors,
the role of the electronic correlation in the superconductivity has been
a subject of intensive investigation.
While conventional BCS superconductors such as Nb and MgB$_{2}$
have good metallic nature of Fermi liquid,\cite{bauer,choi}
unconventional superconductors such as doped cuprate and iron pnictide
show bad metallic behavior due to their
strong electronic correlation.\cite{lee,hauleo}
The correlation issue exists in carbon-based $\pi$-electron
superconductors too.  Ca-doped graphite (CaC$_{6}$)
shows conventional superconductivity with weak
electronic correlation,\cite{csanyl}
whereas Cs$_{3}$C$_{60}$
exhibits unconventional superconductivity
with strong electronic correlation.\cite{capone}

\begin{figure}[b]
\includegraphics[width=9.2cm]{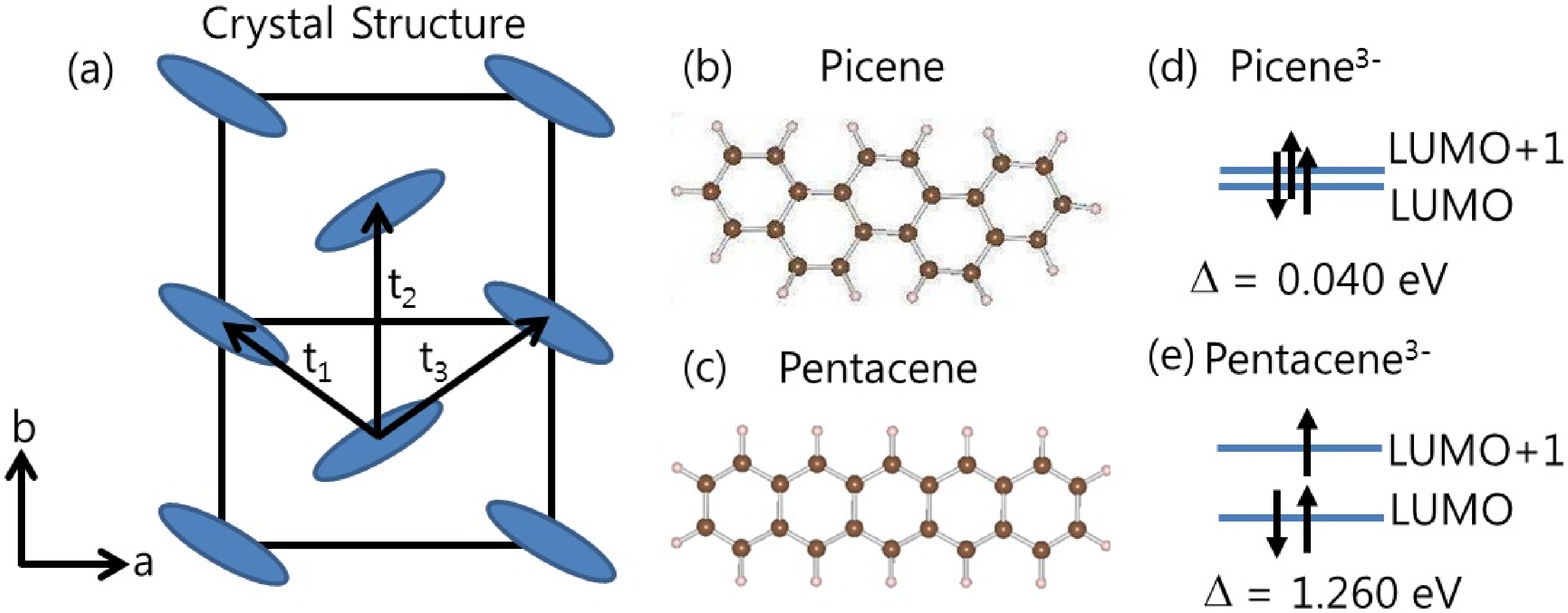}
\caption{(Color online)
(a) Crystal structure of picene and pentacene solids.
Blue ellipses represent picene or pentacene molecular units
shown in (b) and (c).
$t_1$, $t_2$, $t_3$ are hopping amplitudes along each direction.
(b) Molecular structure of picene.
(c) Molecular structure of pentacene.
(d) MO energy levels and occupation of electrons for picene$^{3-}$ solid.
(e) The same for pentacene$^{3-}$ solid.
Note that the energy level splitting ($\Delta$) between LUMO+1 and LUMO
for picene solid ($\Delta=$0.040 eV) is much smaller
than that of pentacene solid ($\Delta=$1.260 eV).
}
\label{fig1}
\end{figure}

New $\pi$-electron superconductors have been recently discovered
in polycyclic aromatic hydrocarbon (PAH)-based molecular solids:
K$_{3}$picene ($T_{c}$=18 K), K$_{3}$coronene ($T_{c}$=15 K),
K$_{3}$phenanthrene ($T_{c}$=5 K),
and K$_{3}$1,2;8,9-dibenzopentacene
($T_{c}$=33 K).\cite{mitsuhashi,kubozono,wang,xue}
These superconductors were also reported to have
strong correlation.\cite{mkim,giovannetti,nomura,ruff}
On the other hand, a similar PAH-based molecular solid,
K-doped pentacene, does not have superconductivity,
but exhibits only the metal-insulator transition (MIT)
behavior.\cite{craciun}
Note that both picene and pentacene are composed of
five benzene rings with slightly different arrangements,
as shown in Fig.~\ref{fig1}.
The different ground states in K-doped picene and K-doped pentacene suggest
that the correlation effects come into play distinctly
between superconducting and non-superconducting
systems.

In this Letter,
in order to resolve the issue of correlation effects
in hydrocarbon-based superconductors,
we have investigated their electronic structures systematically,
employing the dynamical mean-field theory (DMFT).
Based on the ground state electronic structures,
we have constructed the phase diagram of hydrocarbon molecular solids
as functions of doping and relevant energy parameters including
the Coulomb correlation, the Hund coupling, and
the molecular-orbital (MO) energy level splitting.
Our studies reveal that
hydrocarbon-based superconductors belong to multi-band Fermi liquid system,
while non-superconducting K-doped pentacene belongs
to single-band system in the proximity of the MIT.
Further, we have shown that the energy level splitting
between LUMO+1 and LUMO (lowest unoccupied MO)
plays a key role in the superconductivity of
electron-doped hydrocarbon molecular solids.

\begin{table*}[t]
\caption{
Hopping amplitudes $t_{l,m,\textbf{R}}$ (in meV)
determined from the $ab$ $initio$ band structures
of picene and pentacene solids,
where $l$ and $m$ represent the MOs, and
\textbf{R} represents the hopping direction in the lattice.
Subscript 1,2,3 represent the hopping directions as shown in
Fig~\ref{fig1}, and $c$ represents
the direction normal to $ab$ plane.
}
\begin{ruledtabular}
\begin{tabular}{l  c  c  c  c  c  c  c  c  c  c  c}
           ~ & ~t$_{L+1,L,0}$
             & ~t$_{L+1,L+1,1}$
             & ~t$_{L,L,1}$
             & ~t$_{L+1,L+1,2}$
             & ~t$_{L,L,2}$
             & ~t$_{L+1,L+1,3}$
             & ~t$_{L,L,3}$
             & ~t$_{L+1,L+1,c}$
             & ~t$_{L,L,c}$
             & ~t$_{L+1,L,c}$\\
\hline
Picene  & $-$10 & 40 & $-$40 & $-$30 & $-$50 & $-$20 & $-$20 &  0 & 0 & 20\\
Pentacene  & 0 & 60 & 70 & $-$10 & $-$30 & $-$30 & $-$60 &  $-$10 & $-$10 & 0\\
\end{tabular}
\label{table1}
\end{ruledtabular}
\end{table*}

Both picene and pentacene solids have layered crystal structures,
as shown in Fig.~\ref{fig1}(a).
Picene molecule has an arm-chair type structure (Fig.~\ref{fig1}(b)),
while pentacene molecule has a linear arrangement
of benzene rings (Fig.~\ref{fig1}(c)).
The difference in molecular structures produces
the different electronic structures in three electron-doped
picene (picene$^{3-}$) and pentacene (pentacene$^{3-}$).
As shown in Fig.~\ref {fig1}(d) and (e),
the energy level splitting $\Delta$ between LUMO+1 and LUMO
of picene solid ($\Delta=0.04$ eV)
is much smaller than that of pentacene solid ($\Delta$=1.26 eV).
Because the bandwidth of each orbital
in both molecular solids is $\sim$0.25 eV,
picene$^{3-}$ has the occupation of three electrons
on nearly two-fold degenerate orbitals,\cite{mkim,nomura,kosugi1}
whereas pentacene$^{3-}$ has one electron on the single
LUMO+1 orbital that is far separated from the lower LUMO.

The correlation effects in electron-doped hydrocarbon solids are
dealt with by the following two-band Hubbard model Hamiltonian,
\begin{eqnarray}
\label{total}
H&=&H_{0} + H_{I},   \nonumber  \\
 &=& \sum_{l,m,\textbf{R},\sigma}t_{l,m,\textbf{R}}
c_{l,\textbf{R},\sigma}^{\dagger}c_{m,\textbf{0},\sigma}
+ H_{I},
\end{eqnarray}
where $H_{0}$ and $H_{I}$ are non-interacting and interacting Hamiltonians
of doped electrons, respectively.
Here $t_{l,m,\textbf{R}}$ corresponds to hopping from ($0$,$m$) to
($\textbf{R}$,$l$), where $0$ and $\textbf{R}$ represent sites,
$m$ and $l$ represent the MOs.
We have determined $t_{l,m,\textbf{R}}$, using the downfolding scheme
of Kohn-Sham orbitals in the maximally localized Wannier function
(MLWF) basis.\cite{MLWFs,rotation,kosugi2}
Kohn-Sham orbitals were obtained in the generalized gradient
approximation (GGA),
by employing the full-potential augmented plane wave (FLAPW)
band method\cite{FLAPW} implemented in WIEN2k package.\cite{Blaha}
For electron-doped systems, we have utilized the rigid band approximation
due to the absence of experimental crystal structures.
We have employed experimental crystal structures for
undoped picene and pentacene.\cite{de,campbell}
We have confirmed that the structure optimization
in the GGA does not change the original undoped crystal structures much.
The obtained hopping parameters $t_{l,m,\textbf{R}}$ are provided
in Table~\ref{table1}.

$H_{I}$ is given approximately by
\begin{eqnarray}
\label{int}
H_{I}&=&U\sum_{m,\textbf{R}}n_{m,\textbf{R},\uparrow}n_{m,\textbf{R},\downarrow}
+U'\sum_{m>l,\textbf{R},\sigma}n_{m,\textbf{R},\sigma}n_{l,\textbf{R},\bar{\sigma}}
\nonumber \\
&+&(U'-J)\sum_{m>l,\textbf{R},\sigma}n_{m,\textbf{R},\sigma}n_{l,\textbf{R},\sigma},
\end{eqnarray}
where $U$, $U'$ and $J$ are intra-, inter-orbital Coulomb correlation
and Hund interaction parameters, respectively.
We considered here rotationally symmetric interaction, so that
$U_{LUMO+1}\sim U_{LUMO}$ and $U'\sim U-2J$.\cite{nomura}
We have solved the above Hamiltonian by carrying out
the DMFT calculation.\cite{kotliar}
We used the continuous time quantum Monte-Carlo (CTQMC)
method as an impurity solver.\cite{haulet,werner}
We set temperature at T=77 K (=6.67 meV), which is low enough
to observe the MIT in the phase diagram of hydrocarbon solids.\cite{georges}

\begin{figure}[b]
\includegraphics[width=9.0cm]{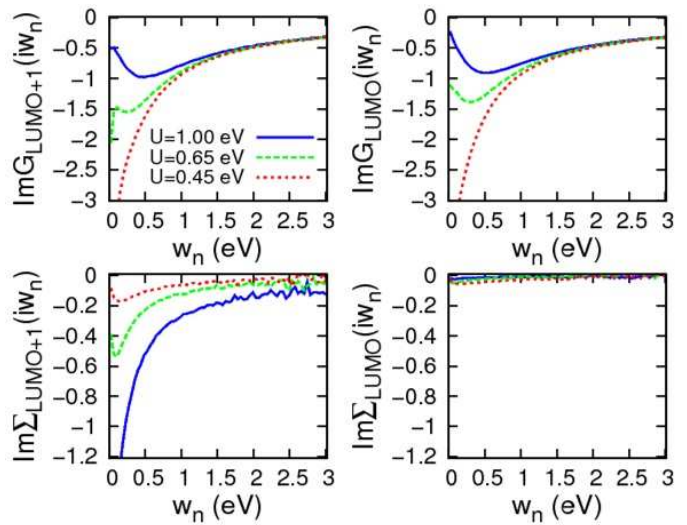}
\caption{(Color online)
Imaginary Green's functions $G(i\omega_n)$ and self energies $\Sigma(i\omega_n)$
of LUMO+1 and LUMO
for picene$^{3-}$ solid ($J$=0.05 eV and $\Delta$=0.04 eV).
Red (dotted), green (dashed), and blue (solid) lines
are for $U$=0.45 eV, $U$=0.65 eV, and $U$=1.00 eV, respectively.
}
\label{fig2}
\end{figure}

Figure~\ref{fig2} shows the imaginary part of
the obtained Green's function $G(i\omega_n)$
and the self-energy $\Sigma(i\omega_n)$ for
LUMO+1 and LUMO of picene$^{3-}$ solid.
For $U=0.45$ eV, Im$G(i\omega_n)$'s of both LUMO+1 and LUMO have finite
values at $\omega_n=0$ ($\omega_n$: Matsubara frequency),
implying that the densities of states (DOSs)
at the Fermi level ($E_F$) are finite for both orbitals.
Also Im$\Sigma(i\omega_n)$'s for both MOs converge
to zero in the zero frequency regime, signifying the Fermi liquid nature.
Note that the magnitude of
Im$\Sigma(i\omega_n)$ is significantly larger for LUMO+1 than for LUMO,
which reflects that the electronic correlation is stronger for LUMO+1.
In fact, $U=0.45$ eV corresponds to the value obtained from the
first-principles calculation for picene$^{3-}$ solid.\cite{nomura}
So these results indicate that K$_{3}$picene solid in its normal state
has two-band Fermi liquid nature with clear orbital-selective band renormalization.

For $U$=0.65 eV, Im$G(i\omega_n)$'s near  $\omega_n=0$
appear to be finite for both LUMO+1 and LUMO,
but that for LUMO has negative slope.
From the derivative of Im$G(i\omega_n)$ near $\omega_n=0$
one can distinguish between metallic and insulating phases
as shown in previous literatures.\cite{kovacik,fuchs}.
We consider that
positive (negative) derivative illustrates the metallic (insulating)
electronic structure.
Accordingly, only the LUMO+1 band has nonzero DOS at $E_F$.
The behaviors of Im$G(i\omega_n)$ and Im$\Sigma(i\omega_n)$
of LUMO+1 for $U$=0.65 eV reflect that the system belongs
to narrow single-band Fermi liquid state.
For $U$=1.0 eV, both Im$G(i\omega_n)$'s are vanishing at $\omega_n=0$,
implying that DOSs at $E_F$ are zero for both LUMO+1 and LUMO.
Im$\Sigma(i\omega_n)$ at $\omega_n=0$ diverges for LUMO+1,
indicating that the system has the Mott insulating state
with hole-orbital disproportionation nature at LUMO+1.
Namely, the hole orbital nature changes from
the mixture of LUMO and LUMO+1 to
the single LUMO+1 (see the inset of Fig.~\ref{fig5}).

\begin{figure}[t]
\includegraphics[width=9.0cm]{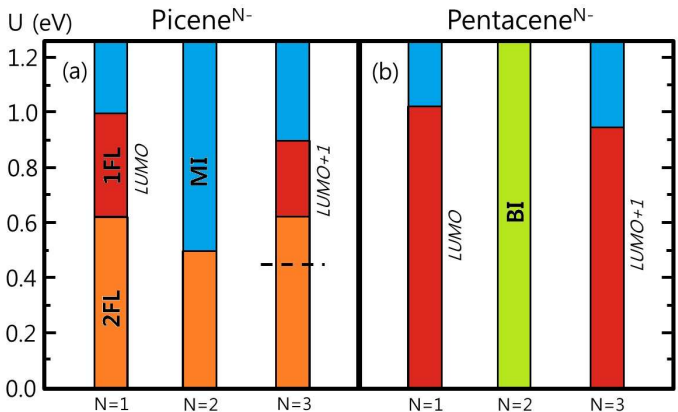}
\caption{(Color online)
(a) Ground states of picene solid ($J$=0.05 eV and $\Delta$=0.04 eV)
with variations of doping (N=1,2, and 3)
and Coulomb interaction ($U$).
(b) Ground states of
pentacene solid ($J$=0.05 eV and $\Delta$=1.26 eV).
1FL, 2FL, BI, and MI stand for one-band Fermi liquid,
two-band Fermi liquid, band insulator, and  Mott insulator,
respectively.
Dashed line for picene$^{3-}$ corresponds to the $U$ value obtained from
the first-principles calculation.\cite{nomura}
}
\label{fig3}
\end{figure}

Figure~\ref{fig3} shows the ground states of
picene and pentacene solids depending on doping $N$ and
the $U$ value.
We have obtained the ground states by following the steps in Fig.~\ref{fig2}.
As mentioned above,
picene$^{3-}$ solid, which has $U=0.45$ eV, belongs to the two-band
Fermi liquid state.\cite{U}
This feature reveals that the superconductivity in
three electron-doped picene, such as K$_{3}$picene and Ca$_{1.5}$picene,
emerges from the multi-band Fermi liquid state.
If we increase $U$ further for picene$^{3-}$, single-band (LUMO+1)
Fermi liquid state is realized at $U$=0.625 eV,
and the Mott insulating state at $U$=0.90 eV.
In the DMFT calculation, Ruff \emph{et al.}\cite{ruff}
used $U$=1.6 eV, which was estimated
from the empirical cavity method for the
screening of electronic correlation.\cite{giovannetti}
This $U$ value is considerably larger than $U=0.45$ eV
obtained from the first-principles calculation.\cite{nomura,U}
That is why Ruff \emph{et al.} obtained the Mott insulating state for K$_{3}$picene
in contrast to our result.
Nevertheless, the occurrence of the MIT for large $U$ is consistent between two.

Similarly to picene$^{3-}$, picene$^{1-}$ solid also
shows the phase transition upon increasing $U$,
from two-band to one-band (LUMO) Fermi liquid state, and then to
the Mott insulating state.
In the case of picene$^{2-}$, due to the finite Hund coupling $J$ (=0.05 eV),
the system for small $U$ remains as two-band Fermi liquid state
with high-spin configuration. With increasing $U$,
the phase transition occurs directly
from two-band Fermi liquid to Mott insulating state.
If one takes into account the non-rigidity of band structures
arising from the hybridization with cations in
K- or Ca-doped picene,
the electronic correlation effect would be further reduced
with respect to the above rigid band case.\cite{kosugi2}
Then K$_{3}$picene and Ca$_{1.5}$picene would have more stable
two-band Fermi liquid nature in their normal states.
If the K doping level is biased from the integer value,
the correlation effect would be not so significant as for the integer doping case.
As a result, one would obtain the stable two-band Fermi liquid state
for the non-integer doping case too.
However, in the case of non-integer doping,
the disorder effect is expected to become important.
In fact, the insulating nature in superconducting K$_x$picene ($x=3.1, 3.5$)
observed above $T_C$ was explained by the granular-metal-like behavior,
which would be attributed to the disorder effect.\cite{teranishi}

\begin{figure}[t]
\includegraphics[width=9.0cm]{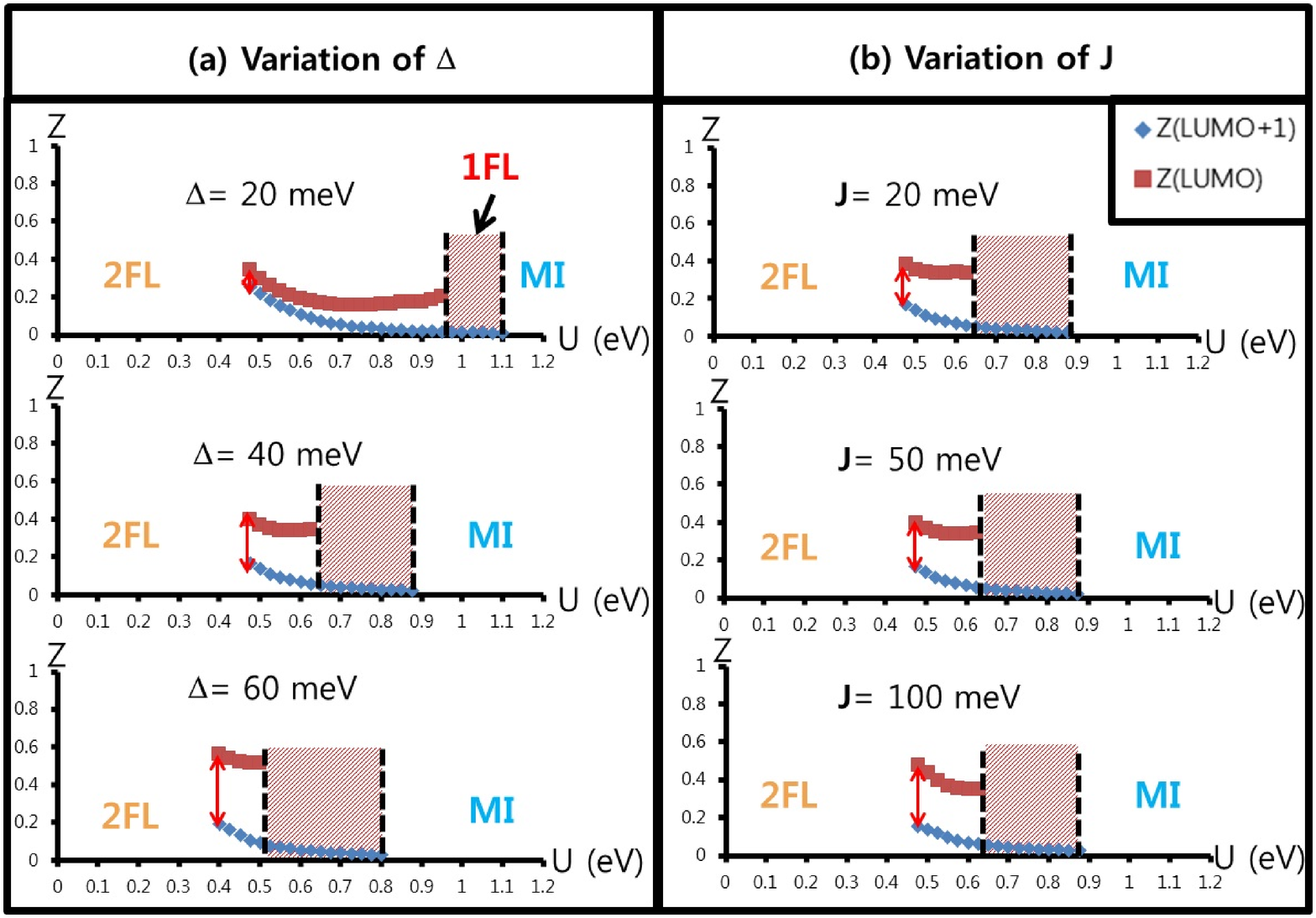}
\caption{(Color online)
(a) Quasi-particle residue $Z$'s
of LUMO+1 and LUMO vs. $U$ for picene$^{3-}$ solid
with variation of $\Delta$ ($J$=50 meV).
(b) The same for picene$^{3-}$ solid
with variation of $J$ ($\Delta$=40 meV).
Red arrow denotes the difference of $Z$'s between LUMO+1 and LUMO.
}
\label{fig4}
\end{figure}

Noteworthy in Fig.~\ref{fig3}(b) is that the ground state
of pentacene$^{3-}$ is very different from that of picene$^{3-}$.
Due to much larger $\Delta$ value in pentacene,
both pentacene$^{3-}$ and pentacene$^{1-}$ solids exhibit the
transition from single-band Fermi liquid to Mott insulator,
like a single-band half-filled system.
On the other hand, pentacene$^{2-}$ has the low-spin state
due to large $\Delta$,
and so only the band insulating state is realized.
The findings in Fig.~\ref{fig3} manifest that
different $\Delta$'s of picene and pentacene
play key roles in producing the different phases.
Also high-spin and low-spin configurations of picene$^{2-}$
and pentacene$^{2-}$ suggest that the Hund coupling $J$
affects the phase diagram.

Figure~\ref{fig4} presents the quasi-particle residue $Z$
for picene$^{3-}$ solid with
the variation of $\Delta$ and $J$.
The quasi-particle residue $Z$ is obtained from
 $Z=(1-\frac{Im\Sigma(iw_{o})}{w_{0}})^{-1}$, where $w_{0}$ is
the lowest Matsubara frequency.\cite{kovacik}
With increasing $\Delta$,
the following features are observed in Fig.~\ref{fig4}(a):
(i) the larger difference of $Z$'s between LUMO+1 and LUMO,
(ii) the wider single-band (LUMO+1) Fermi liquid regime,
and (iii) the reduced critical $U$ for the MIT.
The above observations indicate that just
a small enhancement of $\Delta$ would induce the
single-band Fermi liquid state in picene$^{3-}$ solid,
and thereby suppress the superconductivity.
This sensitive dependence of electronic structure of picene$^{3-}$
upon variation of $\Delta$ explains different experimental
electronic structures of K$_{3}$picene solid depending on
the preparation condition.\cite{ruff,teranishi,mahns1,caputo}
As shown in Fig.~\ref{fig4}(b),
the larger $J$ also yields the larger difference of $Z$'s
between LUMO+1 and LUMO.
But the single-band Fermi liquid regime and
the critical $U$ value for the MIT do not change much.
Lager difference of $Z$'s between two MOs implies
that the inter-orbital fluctuation becomes suppressed.
Therefore Fig.~\ref{fig4} indicates that
the larger $\Delta$ induces the orbital disproportionation,
and the larger $J$ suppresses the inter-orbital fluctuation.\cite{medici}

\begin{figure}[t]
\includegraphics[width=9.0cm]{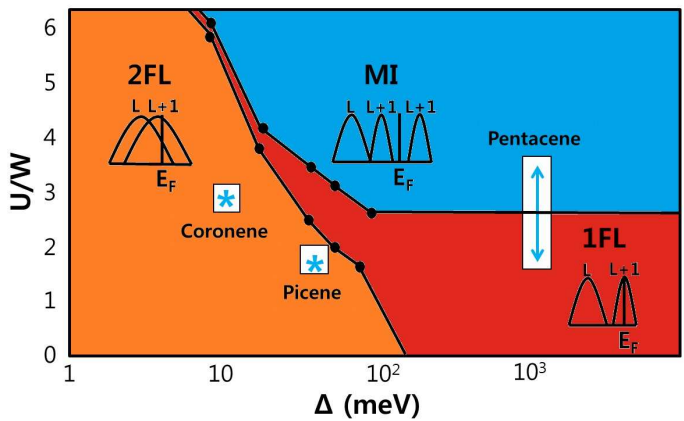}
\caption{(Color online)
Phase diagram of hydrocarbon$^{3-}$ solids
with respect to Coulomb correlation $U$ and MO level splitting $\Delta$.
Positions of picene$^{3-}$, coronene$^{3-}$, and pentacene$^{3-}$ solids
are marked in the phase diagram.
Inset diagram represents the schematic DOS of each phase.
In the MI regime, the hole-orbital disproportionation nature
is realized.
}
\label{fig5}
\end{figure}

Figure~\ref{fig5} presents the overall phase diagram
of hydrocarbon$^{3-}$ solids as functions of $U$ and $\Delta$.
The ground state of picene$^{3-}$ solid is
two-band Fermi liquid state.
In contrast, the  ground state of pentacene$^{3-}$ solid
is single-band Fermi liquid or Mott insulator
depending on $U$ value.\cite{Upentacene}
This difference is the reason why picene$^{3-}$ is superconducting,
while pentacene$^{3-}$ is non-superconducting.
The ground states of K$_{3}$coronene and K$_{3}$phenanthrene
in their normal states are also two-band Fermi liquid states,
with three doped electrons occupying nearly degenerate LUMO and LUMO+1.
On the other hand, 1,2;8,9-dibenzopentacene has large energy splitting
between LUMO and LUMO+1, but nearly degenerate LUMO+1 and LUMO+2.\cite{mahns2}
Hence K$_{3}$1,2;8,9-dibenzopentacene
has one electron on the nearly degenerate LUMO+1 and LUMO+2.
This situation is similar to that of picene$^{1-}$ in Fig.~\ref{fig3}(a).

Dopant-induced structural deformation and hybridization are to be
reflected in the variation of $\Delta$ and $U/W$ in Fig.\ref{fig5}.
For example, the total bandwidth of K$_{3}$picene is enhanced
with respect to that of undoped picene, from $\sim$0.35 eV to $\sim$0.60 eV.\cite{kosugi2}
As a result, the position of K$_{3}$picene would be further lowered than that of picene$^{3-}$
in the phase diagram of Fig.~\ref{fig5}.\cite{Uanddelta}
Therefore the phase diagram of Fig.~\ref{fig5} demonstrates
that the superconductivity in hydrocarbon-based superconductors commonly
emerge from the two-band Fermi liquid state with good metallic nature.
This feature supports the conventional phonon-mediated BCS mechanism
rather than other exotic mechanisms for the superconductivity
of hydrocarbon-based molecular solids,\cite{kato,casula1,casula2,subedi}
despite that they have $U/W$ larger than one.
The underlying superconducting mechanism in hydrocarbon-based superconductors
needs further investigation, but it is evident that the superconductivity
would be boosted up by the enhanced DOS at $E_F$ due to their
two-band Fermi liquid nature.\cite{kato}
Our phase diagram indicates that small $U/W$ and $\Delta$ are key factors
for the emergence of superconductivity in hydrocarbon-based molecular solids.
Thus the superconductivity is preferentially to be searched for
in closely packed hydrocarbon-based molecular solids that have small $U/W$ and $\Delta$.


In conclusion, we have investigated the electronic structures
of electron-doped hydrocarbon solids based on the DMFT calculations,
and constructed the phase diagram with respect to
Coulomb correlation $U$, doped electrons $N$, Hund coupling $J$,
and MO energy level splitting $\Delta$.
We have shown that the superconductivity in hydrocarbon solids
commonly emerges from the two-band Fermi liquid state.
This is in contrast to the case of non-superconducting pentacene,
which has the effective single-band Fermi liquid state in
the proximity of the MIT.
The size of MO energy level splitting plays an important  role
in determining the ground states of hydrocarbon solids.
Our results demonstrate that multi-band nature in hydrocarbon solids
is essential to boost the superconductivity
through the enhanced DOS at $E_F$.
It is thus suggested that higher $T_C$ superconductors
need to be searched for in
more closely packed
molecular solids with multi-band nature at $E_F$.

\begin{acknowledgments}
Discussions with Ki-Seok Kim and Beom Hyun Kim are gratefully acknowledged.
This work was supported by the National Research Foundation of Korea
(Grant Nos. 2009-0079947, 2010-0006484, R32-2008-000-10180-0).
\end{acknowledgments}


\begin{thebibliography}{99}
\bibitem{rotation} MLWF provides lower and upper Wannier functions,
     which are $\frac{1}{\sqrt{2}}(\psi_{LUMO+1}+\psi_{LUMO})$ and
     $\frac{1}{\sqrt{2}}(\psi_{LUMO+1}-\psi_{LUMO})$.\cite{kosugi1,kosugi2}
	We applied the unitary transform to obtain Hamiltonian $H_{0}$
     with LUMO+1 and LUMO basis.
\bibitem{U} Here $U$ is corresponding to an effective Coulomb
	interaction that is given by $\bar{U}$-$\bar{V}$
	in Ref.\cite{nomura}, where $\bar{U}$ and $\bar{V}$ represent on-site
    	and nearest neighboring Coulomb interactions, respectively.
\bibitem{Upentacene} First-principle result of $U$ for pentacene$^{3-}$
	is not available. The range of  $U$ in Fig.~\ref{fig5} corresponds to
	that in-between the first-principles (Ref.\cite{nomura})
	and the empirically calculated $U$ (Ref.\cite{mkim}) for picene$^{3-}$.
\bibitem{Uanddelta}
     There are several metastable
     structures for K$_{3}$picene.\cite{ruff,kosugi2,teranishi,mahns1,caputo}
     Determination of $\Delta$ and $U/W$ for each metastable structure
	 would be an interesting future problem.
\end{thebibliography}
\end{document}